\documentclass{ws-p8-50x6-00}

\usepackage{amsmath,amssymb}
\usepackage[english]{babel}

\newcommand{\LT}   {\left}
\newcommand{\RT}   {\right}

\newcommand{\MeV}  {{\text{~MeV}}}
\newcommand{\GeV}  {{\text{~GeV}}}
\newcommand{\EtAl} {\emph{et~al}}
\newcommand{\CN}   {{\tilde\chi^{\pm,0}}}

\begin{document}

\title{Bounds on ``charginos nearly degenerate with the lightest
  neutralino'' mass from precision measurements}

\author{M.~Maltoni}

\address{
  Instituto de F\'{\i}sica Corpuscular~--~C.S.I.C., \\
  Departament de F\'{\i}sica Te\`orica, Universitat de Val\`encia, \\
  Edificio Institutos de Paterna, Apt.~2085, E-46071 Valencia, Spain \\
  INFN, Sezione di Ferrara, Via Paradiso 12, I-44100 Ferrara, Italy \\
  E-mail: maltoni@ific.uv.es}

\maketitle

\abstracts {
  Though LEP~II direct searches still cannot exclude a chargino nearly
  degenerate with the lightest neutralino if its mass is only slightly above
  half of the $Z$ boson mass and the sneutrino is light, it can be excluded
  indirectly analyzing precision data. In this particular limit simple
  analytical formulas for oblique electroweak radiative corrections are
  presented.}

\section{Introduction}

According to the latest searches performed at LEP~II at center-of-mass
energies up to $189\GeV$, the present bounds on chargino mass are
$m_{\tilde\chi^\pm}\gtrsim 90\GeV$ for the higgsino-dominated case (or when
the sneutrino is heavy) and $m_{\tilde\chi^\pm}\gtrsim 80\GeV$ in the
wino-dominated light-sneutrino scenario.~\cite{ALEPH99,DELPHI00} However, when
the lightest chargino and neutralino (the latter being the LSP) are almost
degenerate in mass, the charged decay products of the light chargino are very
soft, and the above quoted bounds are no longer valid. A special search for
such light charginos has been performed recently by the DELPHI collaboration,
and the case of $\Delta M^\pm\equiv m_{\tilde\chi_1^\pm} - m_{\tilde\chi_1^0}
\lesssim 100\MeV$ is now excluded.~\cite{ALEPH99,DELPHI00} In the region
$\Delta M^\pm\sim 1\GeV$ the analysis of the Initial State Radiation (ISR) can
be used to put a limit on the chargino mass,~\cite{DELPHI00} but in the case
of wino domination with a light sneutrino this technique fails and charginos
as light as $45\GeV$ (this bound coming from the measurements of $Z$ decays at
LEP~I and SLC) are still allowed. The case of almost degenerate chargino and
neutralino can be naturally realized in SUSY and the possibilities to find
such particles are discussed in literature.~\cite{Gunion99}

In this talk we investigate the radiative corrections to the electroweak
precision measurements generated by such almost degenerate
particles.~\cite{Maltoni99,MaltoniPHD} When their masses are close to $m_Z/2$
one-loop contributions are large and they spoil the perfect description of
experimental data by the Standard Model. Due to the decoupling property of
SUSY models, when $m_\CN \gg m_Z$ the radiative corrections are power
suppressed.

\section{Discussion}

The contributions of new physics to the electroweak precision data through
oblique corrections can be conveniently parameterized in terms of the three
functions $V_m$, $V_A$ and $V_R$.~\cite{NORV99} In the simplest supersymmetric
extensions of the Standard Model the chargino-neutralino sector is defined by
the numerical values of the four parameters $M_1$, $M_2$, $\mu$ and
$\tan\beta$, and the case of nearly degenerate lightest chargino and
neutralino naturally arise when:
\begin{itemize}
  \item $M_2 \gg \mu$: in this limit the particles of interest form an $SU(2)$
    doublet of Dirac fermions, whose wave functions are dominated by
    \emph{higgsinos}, and their contribution to the $V_i$ functions
    is:~\cite{Maltoni99}
    \begin{align}
	\begin{split} 
	    \delta^{\tilde h} V_m &= \frac{16}{9} \bigg[
	    \LT( \frac{1}{2} - s^2 + s^4 \RT) \LT( 1 + 2\chi \RT) F(\chi) \\
	    & \hspace{1cm} - \LT( \frac{1}{2} - s^2 \RT)
	    \LT( 1 + 2\frac{\chi}{c^2} \RT) F\LT( \frac{\chi}{c^2} \RT)
	    - \frac{s^4}{3} \bigg],
	\end{split} \\[2mm]
	\delta^{\tilde h} V_A &= \frac{16}{9} \LT( \frac{1}{2} - s^2 + s^4 \RT)
	  \LT[ \frac{12 \chi^2 F(\chi) - 2\chi - 1}{4\chi - 1} \RT], \\[2mm]
	\delta^{\tilde h} V_R &= \frac{16}{9} c^2 s^2
	  \LT[ \LT( 1 + 2\chi \RT) F(\chi) - \frac{1}{3} \RT],
    \end{align}
    where $\chi \equiv (m_\CN / m_Z)^2$, the function $F$ is defined in App.~B
    of Ref.~\cite{NORV99}, and $s^2$ ($c^2$) is the sine (cosine) squared of
    the electroweak mixing angle;

  \item\label{case_b} $\mu \gg M_2$: in this case we get an $SU(2)$ triplet
    of Majorana fermions, with the wave functions dominated by \emph{winos},
    and the expressions for the corrections to $V_i$ are:~\cite{Maltoni99}
    \begin{align}
	\begin{split}
	    \delta^{\tilde w} V_m &= \frac{16}{9} \bigg[
	    c^4 \LT( 1 + 2\chi \RT) F(\chi) \\
	    & \hspace{1cm} - \LT( 1 - 2s^2 \RT) \LT( 1 + 2\frac{\chi}{c^2} \RT)
	    F\LT( \frac{\chi}{c^2} \RT) - \frac{s^4}{3} \bigg],
	\end{split} \\[2mm]
	\delta^{\tilde w} V_A &= \frac{16}{9} c^4
	  \LT[ \frac{12 \chi^2 F(\chi) - 2\chi - 1}{4\chi - 1} \RT], \\[2mm]
	\delta^{\tilde w} V_R &= \frac{16}{9} c^2 s^2
	  \LT[ \LT( 1+2\chi \RT) F(\chi) - \frac{1}{3} \RT].
    \end{align}
\end{itemize}

\begin{figure}[t] \centering
    \epsfxsize=0.9\textwidth \epsfbox{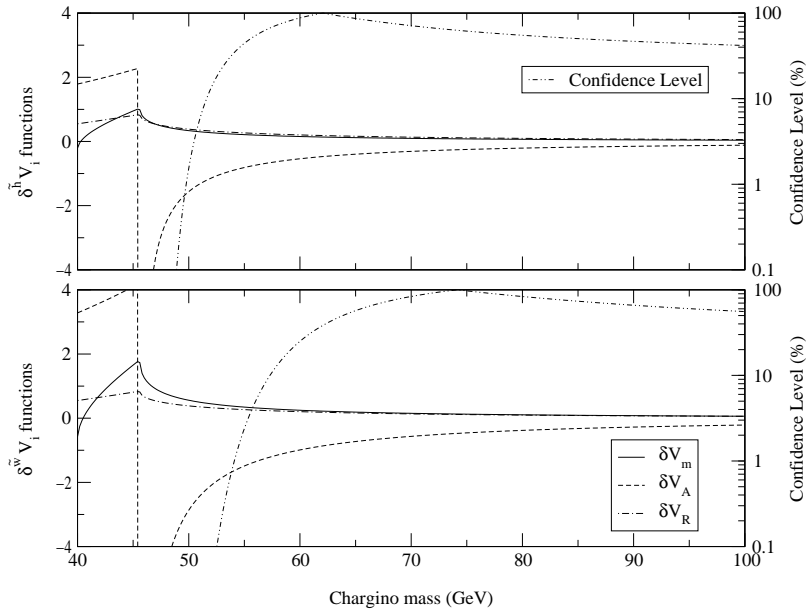}
    \caption{\label{fig:10}
      Dependence of the $\delta V_i$ functions (left Y-axes) and of the
      Confidence Levels (right Y-axes) on the light chargino-neutralino mass
      $m_\CN$, in the limits $M_2 \gg \mu$ (higgsino-dominated case, upper
      panel) and $\mu \gg M_2$ (wino-dominated case, lower panel).}
\end{figure}

The $\chi^2$ for the new physics contributions to $V_i$ was computed using the
computer program \texttt{LEPTOP},~\cite{NORV99,NOV95} and the corresponding
Confidence Level (together with the numerical value of the $\delta V_i$
functions) is plotted in Fig.~\ref{fig:10} against the chargino-neutralino
mass $m_\CN$. We see that at $95\%$ C.L.\ the bounds $m_\CN \gtrsim 51\GeV$
(higgsino-dominated case) and $m_\CN \gtrsim 56\GeV$ (wino-dominated case)
should be satisfied. Note that the main contribution to $\chi^2$ comes from
$\delta V_A$, which is singular at $m_\CN = m_Z/2$. This singularity is not
physical and our formulas are valid only for $2 m_\CN\gtrsim m_Z + \Gamma_Z$;
however, the existence of $\chi^\pm$ with a mass closer to $m_Z/2$ will change
Z-boson Breit-Wigner curve, therefore it is also not
allowed.~\cite{Barbieri92}

\section{Conclusions}

Let us briefly discuss the contributions of other SUSY particles to the $V_i$
functions. In the considered limits the remaining charginos and neutralinos
are very heavy, so they simply decouple. The contributions of the three
generations of sleptons (with masses larger than $90\GeV$) and of the first
two generations of squarks (with masses larger than $200\GeV$ to satisfy
Tevatron bounds) into $V_A$ are smaller than $0.1$, so they can safely be
neglected. Concerning the contributions of the third generation squarks,
although enhanced by the large top-bottom mass difference they are almost
universal,~\cite{Gaidaenko98} so compensating the negative value of $V_A$ will
generate large positive terms into $V_R$ and $V_m$ and the overall $\chi^2$
will not be better. Finally, according to Ref.~\cite{Chankowski94} the
contribution to radiative corrections of the whole MSSM Higgs sector equals
with very good accuracy that of a single SM higgs with the same mass as the
lightest MSSM neutral higgs, so it is already accounted for in our analysis.

Let us remark that in the case of wino domination with a light sneutrino,
which occurs naturally in anomaly-mediated SUSY breaking
scenarios,~\cite{Gunion99} the analysis of the ISR fails~\cite{DELPHI00} and
the bound $m_\CN\gtrsim 56\GeV$ from precision measurements is presently the
strongest constraint which can be imposed on the chargino-neutralino mass.

\section*{Acknowledgments}

I wish to thank my collaborator M.I.~Vysotsky. I am also grateful to
A.N.~Rozanov for evaluating the C.L.\ shown in Fig.~\ref{fig:10} with the
program \texttt{LEPTOP}. This work was supported by DGICYT under grant
PB98-0693 and by the TMR network grant ERBFMRX-CT96-0090 of the European
Union.

\end{document}